\documentstyle[preprint,revtex]{aps}

\begin{document}
\thispagestyle{empty}

% TITLE PAGE

\begin{center}
\hfill{IP-ASTP-13-94}\\
\hfill{July 1994}

\vspace{1 cm}

\begin{title}
Random and Correlated Phases of Primordial Gravitaional Waves
\end{title}
\vspace{1 cm}

\author{Kin-Wang Ng and A.~D.~Speliotopoulos}
\vspace{0.5 cm}

\begin{instit}
Institute of Physics, Academia Sinica\\ Taipei, Taiwan 115, R.O.C.
\end{instit}
\end{center}
\vspace{0.5 cm}

\begin{abstract}
\vspace{1 cm}

\noindent The phases of primordial gravity waves is analysed in
detail within a quantum mechanical context following the formalism
developed by Grishchuk and Sidorov. It is found that for physically
relevant wavelengths both the phase of each individual mode
and the phase {\it difference\/} between modes are randomly
distributed. The phase {\it sum\/} between modes with oppositely
directed wave-vectors, however, is not random and takes on a definite
value with no rms fluctuation. The conventional point of view that
primordial gravity waves appear after inflation as a classical,
random stochastic background is also addressed.
\end{abstract}
\newpage

It is well known that gravitational waves (GW) can be produced in
inflationary cosmologies $\cite{Star}$. They arise from quantum
fluctuations in the gravitaional field on a curved de Sitter
space-time. The generated perturbations are then often treated as a
Gaussian noise with randomly distributed phases. During de Sitter
expansion, it is believed that these fluctuations are then
stretched ``out'' of the horizon where they ``freeze'' and remain
constant until much later when they re-enter the horizon
and appear as classical GW. These GW then form a classical
stochastic background which fills the universe. It has, however,
also been argued that the quantum-mechanical generation of
perturbations, in constrast to a classical, stochastic perturbation
spectrum, actually possesses very specific statistical
properties. Specifically, it has been argued that the phases of all modes
of perturbations are essentially constant and fixed. Moreover,
strong quantum correlations cause classical standing GW to form
leading to important cosmological consequences $\cite{Gris}$.

In this paper we shall present a detailed analysis of the statistical
properties of the phases of relic GW from a quantum mechanical
viewpoint. As the generation of these GW are fundamentally quantum
mechanical in nature we shall
thus be able to comment on the view that these GW subsequently form a
classical, stochastic background. Our results show that the phase of each
individual mode is random. More importantly, in all but one
case, there is also the absence of any correlation
between the phases of two different modes. Namely, both the phase sum
and phase difference between any two aribitrary modes vanish. There
exists, however, one exceptional case, that of two modes with
oppositely directed wave-vectors, for which the {\it phase sum\/} of
the two modes are highly-correlated. This is a manifestation of phase
locking between modes which also occurs in quantum optics
$\cite{BarPegg}$. The phase difference between these two modes are
still completely uncorrelated and random, however. In many instances,
therefore, the relic GW may be {\it approximated\/} as a
``classical'' stochastic background. Nevertheless, the phase-sum
locking asserts that the background is fundamentally {\it quantum
mechanical\/} in nature. Based on these results, we shall comment on
the usual classical treatment of relic GW.

In our analysis we shall follow the framework developed by Grishchuk
and Sidorov $\cite{Gris}$. But instead of using `standing wave'
operators which decompose the two-mode squeeze operator into the product
of two single mode operators, we shall stay with the original
`traveling wave' operators where the physics is more
straightforwardly seen. We warn the reader,
however, that while the number operator for each mode is invariant
under either choice, the phase operator is not and our definition of
the phase operator differs from that given in $\cite{Gris}$

The quantized graviton field operator $h_{ij}$ may be written as
\begin{equation}
h_{ij}(\eta, \vec x) = C\sum_{\vec q}\sum^2_{s=1} p^s_{ij}(\vec q)
        \left[a^s_{\vec q}(\eta)e^{i\vec q\cdot\vec x}+{a^s_{\vec
        q}}^\dagger(\eta)e^{-i\vec q\cdot\vec x}\right]\>,
\label{e1}
\end{equation}
where $C$ is an overall constant whose value is not imporant for our
purposes, $p^s_{ij}(\vec q)$ are the polarization tensors and $s$ labels
the two polarization states of the wave. $a_{\vec q}^s(\eta)$ and
${a_{\vec q}^s}^\dagger(\eta)$ are raising and lowering operators in
the Heisenberg representation and with their evolution governed by the
hamiltonian
\begin{equation}
H = \sum_{\vec q}\sum_{s=1}^2
        \left\{
                q{a^s_{\vec q}}^\dagger(\eta)a^s_{\vec q}(\eta) +
                q{a^s_{-\vec q}}^\dagger(\eta)a^s_{-\vec q}(\eta) +
               2\sigma(\eta)
                \left[
                {a^s_{\vec q}}^\dagger(\eta){a^s_{-\vec q}}^\dagger(\eta)-
                a^s_{\vec q}(\eta)a^s_{-\vec q}(\eta)
                \right]
        \right\}
\label{e2}
\end{equation}
where $\sigma(\eta)=iR'/2R$. Here $R$ is the cosmic scale factor and
the prime denotes derivative with
respect to $\eta$.

Notice that the graviton field is coupled to the background metric,
and thus to the expansion of the universe, through a conformal time
varying {\it quadratic\/} interacting hamiltonian. This
hamitonian may be diagonalized and the Heisenberg evolution equations
\begin{equation}
ida^s_{\vec q}/d\eta = [a^s_{\vec q},H]\>,\qquad id{a^s_{\vec
q}}^\dagger/d\eta = -[{a^s_{\vec q}}^\dagger,H]\>,
\end{equation}
solved exactly using a Bogolubov transformation. In
fact, Grishchuk and Sidorov make the time dependent Bogolubov
transformation: $a^s_{\vec q}(\eta)\to u_q(\eta)a^s_{\vec q}(\eta_0) +
v_q(\eta){a^s_{-\vec q}}^\dagger(\eta_0)$, ${a^s_{\vec
q}}^\dagger(\eta) \to \bar u_q(\eta) {a^s_{\vec q}}^\dagger(\eta_0) +
\bar v_q(\eta)a^s_{-\vec q}(\eta_0)$ where $\eta_0$ is some initial
conformal time. Requiring this to be a canonical transformation
restricts
\begin{equation}
\vert u_q(\eta)\vert^2 - \vert v_q(\eta)\vert^2=1,
\label{e3}
\end{equation}
for all $\eta$. Then, noting that in the
presence of inflation $R'/R\to0$ as $\eta\to-\infty$, they choose as the
initial conditions for the Bogolubov coefficients to be
$u_q(\eta_0)=1$ and $v_q(\eta_0)=0$ where $\eta_0$ is some initial
time in the far past.

While this is certainly a valid approximation, one eventually finds
that it is more useful to write
\begin{equation}
u_q(\eta) = e^{i\epsilon_q}\cosh r_q\>, \qquad
v_q(\eta) = e^{-i(\epsilon_q-2\phi_q)}\sinh r_q\>,
\label{e4}
\end{equation}
which explicitly satisfies eq.~$(\ref{e3})$.
The functions $r_q(\eta)$, $\epsilon_q(\eta)$
and $\phi_q(\eta)$ are called the squeeze parameter, rotation angle, and
squeeze angle, respectively. If one then uses Grishchuk's
initial conditions, one finds that $r_q(\eta_0) =0$ and
$\epsilon_q(\eta_0)=0$. The initial condition for $\phi_q$ is
{\it undetermineable}, however, and it is this phase
which will play a crucial role in determining the phase of the GW. A
more careful analysis must there be done to establish the correct
initial condition for it.

To do so, let us begin by defining
\begin{eqnarray}
\alpha^s_{\vec q}(\eta) &=& u_q(\eta)a^s_{\vec q}(\eta_0) +
                          v_q(\eta){a^s_{-\vec q}}^\dagger(\eta_0)\>,
\nonumber \\
{\alpha^s_{\vec q}}^\dagger(\eta) &=& \bar u_q(\eta)
                        {a^s_{\vec q}}^\dagger(\eta_0) +
                        \bar v_q(\eta)a^s_{-\vec q}(\eta_0)\>,
\label{e5}
\end{eqnarray}
where $a^s_{\vec q}(\eta_0)$ and ${a^s_{\vec q}}^\dagger(\eta_0)$ are
evaluated at some initial time $\eta_0$. Then from the Heisenberg
evolution equations, $u_q$ and $v_q$ evolve as
\begin{eqnarray}
\frac{du_q}{d\eta} &=& -iqu_q + \frac{R'}{R}\bar v_q\>,
\nonumber \\
\frac{dv_q}{d\eta} &=& -iqv_q + \frac{R'}{R}\bar u_q\>.
\label{e6}
\end{eqnarray}
The initial conditions for $u_q$ and $v_q$ are given at $\eta_0$ and
are fixed by requiring $\alpha^s_{\vec q}(\eta_0)$ and
${\alpha^s_{\vec q}}^\dagger(\eta_0)$ to diagonalize $H$ {\it at this
time}:
\begin{equation}
H(\eta_0) = \sum_{\vec q}\sum_{s=1}^2
qe(q)\left[{\alpha^s_{\vec q}}^\dagger(\eta_0)\alpha^s_{\vec
q}(\eta_0)+{\alpha^s_{-\vec q}}^\dagger(\eta_0)\alpha^s_{-\vec
q}(\eta_0)\right],
\label{e7}
\end{equation}
since it is only in this way that a ground state for the system can
be defined. From the standard Bogolubov analysis, we find that
\begin{equation}
u_q(\eta_0) = ie^{i\theta}\sqrt{\frac{1+e}{2e}}\qquad, \qquad
v_q(\eta_0) = -e^{i\theta}\sqrt{\frac{1-e}{2e}},
\label{e8}
\end{equation}
and
\begin{equation}
e(q) = \sqrt{1-\left(\frac{R'(\eta_0)}{qR(\eta_0)}\right)^2}.
\label{e9}
\end{equation}
Here $\theta$ is an arbitrary phase for each mode $q$. We fix it by
requiring that $u_q(\eta_0)\to1$ if the interaction term is turned
off: $\vert\sigma\vert\to0$ so that $\theta=-\pi/2$ for all q.
Notice also that because $e(q)$ must be real
number, the initial time $\eta_0$ must be chosen such that
$2\vert\sigma(\eta_0)\vert<q$ for all $\vec q$. Usually $\eta_0$ is
chosen to be at such an early time that
$\vert\sigma(\eta_0)\vert\approx0$ and this is not a problem. From
Eq.~$(9)$, $\cosh r_q(\eta_0) =\sqrt{(1+e)/2e}$, and $\epsilon_q(\eta_0)=0$.
These argee with Grishchuk's initial conditions in the limit $e(q)\to
1$. The initial condition for $\phi_q(\eta_0) = -\pi/4$ can now be
determined, however.

As was first noticed by $\cite{Gris}$, the above system is
equivalent to what are called squeezed states in quantum optics.
Namely, one notices that the tranformed operators are related to the
original operators through a unitary transformation:
\begin{eqnarray}
\alpha^s_{\vec q}(\eta) &=&
        {\cal R}^s_{\vec q}(\eta){\cal S}^s_{\vec q}(\eta)
        a^s_{\vec q}(\eta_0)
        {{\cal S}^s_{\vec q}}^\dagger(\eta)
        {{\cal R}^s_{\vec q}}^\dagger(\eta)\>,
\nonumber \\
{\alpha^s_{\vec q}}^\dagger(\eta) &=&
        {\cal R}^s_{\vec q}(\eta){\cal S}^s_{\vec q}(\eta)
        {a^s_{\vec q}}^\dagger(\eta_0)
        {{\cal S}^s_{\vec q}}^\dagger(\eta)
        {{\cal R}^s_{\vec q}}^\dagger(\eta) \>,
\label{e10}
\end{eqnarray}
where
\begin{equation}
{\cal R}^s_{\vec q}(\eta)=\exp\left\{-i\epsilon_q(\eta)
                \left(
                {a^s_{\vec q}}^\dagger(\eta_0) a^s_{\vec q}(\eta_0)
                +
                {a^s_{-\vec q}}^\dagger(\eta_0) a^s_{-\vec q}(\eta_0)
                \right)
                \right\}\>,
\label{e11}
\end{equation}
is the two mode rotation operator while
\begin{equation}
{\cal S}^s_{\vec q}(\eta) = \exp\left\{r_q(\eta)
                \left(
                e^{-2i\phi_q(\eta)}a^s_{\vec q}(\eta_0)a^s_{-\vec q}(\eta_0)
                -
                e^{2i\phi_q(\eta)}{a^s_{\vec q}}^\dagger(\eta_0)
                {a^s_{-\vec q}}^\dagger(\eta_0)
                \right)
                \right\}
\label{e12}
\end{equation}
is the two mode squeeze operator.

It is currently held that quantum fluctuations in the gravitational
field will be amplified as the universe undergoes a fast expansion.
(See, for example $\cite{BirDav}$ for a
complete description of particle creation in the universe). This can
be seen explicitly by looking at
\begin{equation}
\langle0\vert n^s_{\vec q} \vert0\rangle = \langle0\vert
{\alpha^s_{\vec q}}^\dagger(\eta)\alpha^s_{\vec
q}(\eta) \vert0\rangle = \vert v_q(\eta)\vert^2 = \sinh^2 r_q(\eta)\>,
\label{e13}
\end{equation}
for any mode $\vec q$. It is then argued that at the present time
these primordial, quantum fluctuations will appear as classical GW
with a classical stochastic destribution of random phases.
This, however, has never been explicitly established.

It is known $\cite{CarNieto}$ that the number operator and the
phase operator
\begin{equation}
\exp\{i\varphi^s_{\vec k}(\eta)\} \equiv
\left(I+n^s_{\vec k}(\eta)\right)^{-1/2}\alpha_{\vec k}^s(\eta)\>,\quad
\exp\{-i\varphi^s_{\vec k}(\eta)\} \equiv
{\alpha_{\vec k}^s\dagger(\eta)}\left(I+n^s_{\vec k}(\eta)\right)^{-1/2},
\label{e14}
\end{equation}
for the graviton field do not commute. The phase itself is not an
hermitian operator, however, and suffers from a problem with
multiplicity. We shall instead have to work with the operators
\begin{eqnarray}
\cos\varphi_{\vec k}^s(\eta) &\equiv& \frac{1}{2} \left(\exp\{i\varphi_{\vec
k}^s(\eta)\} + \exp\{-i\varphi_{\vec k}^s(\eta)\}\right)\>,
\nonumber \\
\sin\varphi_{\vec k}^s(\eta) &=& \frac{1}{2i} \left(\exp\{i\varphi_{\vec
k}^s(\eta)\} - \exp\{-i\varphi_{\vec k}^s(\eta)\}\right)\>,
\label{e15}
\end{eqnarray}
which are hermitian, and thus physical observables. One then finds
that
\begin{equation}
[n^s_{\vec k}, \cos\varphi_{\vec k}^s] = -i\sin\varphi_{\vec k}^s
\>, \qquad
[n^s_{\vec k}, \sin\varphi_{\vec k}^s] = i\cos\varphi_{\vec
k}^s\>,
\label{e16}
\end{equation}
with the corresponding uncertainty relations
\begin{equation}
\Delta n^s_{\vec k} e\cos\varphi_{\vec k}^s\ge
\frac{1}{2}\vert\langle\sin\varphi_{\vec k}^s\rangle\vert\>, \qquad
\Delta n^s_{\vec k} e\sin\varphi_{\vec k}^s\ge
\frac{1}{2}\vert\langle\cos\varphi_{\vec k}^s\rangle\vert\>,
\label{e17}
\end{equation}
where $\Delta A \equiv \sqrt{\langle A^2\rangle-\langle A\rangle^2}$
is the rms fluctuation of the operator. The classical limit for
massless particles, therefore, involves not only looking at the
intensity of the radiation, but also at its phase.

Using eq.~$(\ref{e4})$, the rms fluctuation in the number operator
for each mode is
\begin{equation}
\Delta n_{\vec k}^s = \vert u_k\vert\vert v_k\vert = \sqrt 2\cosh
r_k(\eta)\sinh r_k(\eta)
\label{e18}
\end{equation}
so that $\Delta n_{\vec k}^s /\langle n_{\vec k}^s\rangle = \sqrt 2/\tanh
r_k$. Now, it is usually stated $\cite{Gris}$ that although
$\langle0\vert h_{ij}\vert0\rangle=0$, because $\langle0\vert n_{\vec
k}^s\vert0\rangle\ne0$, for $\vec k$ mode with a large squeeze
parameter one can still view each mode of the GW as a classical wave
with amplitude $A_{\vec k} = \langle0\vert n_{\vec
k}^s\vert0\rangle^{1/2}$. One can certainly make this interpretation,
but it does give the impression that the classical wave will have a
well defined, definite amplitude. From eq.~$(\ref{e18})$, we see
that this need not be the case. In fact, in inflationary cosmologies
$\cite{Gris}$, modes just entering the horizon at
the present day would have a $r_k\sim 120$ giving
$\Delta n_{\vec k}^s \sim \langle n_{\vec k}^s\rangle$.
The fluctuation of
the amplitude of this classical wave will be on the order of its
amplitude itself. In this case, we {\it cannot\/} characterize the
resultant `classical' wave as having any definite amplitude.

Let us now consider the phases of these GW. We begin by calculating
the average phase of each mode
\begin{equation}
\langle0\vert\cos\varphi_{\vec q}^s(\eta)\vert0\rangle
        = \langle0\vert {\cal R}^s_{\vec q}(\eta){\cal S}^s_{\vec
q}(\eta)\cos\varphi_{\vec k}^s(\eta_0){{\cal S}^s_{\vec
q}}^\dagger(\eta) {{\cal R}^s_{\vec q}}^\dagger(\eta)\vert0\rangle\>,
\label{e19}
\end{equation}
where we have used eq.~$(\ref{e10})$. Since ${{\cal R}^s_{\vec
q}}^\dagger\vert0\rangle =\vert0\rangle $, and by using the following
factorization $\cite{ShuCaves}$,
\begin{eqnarray}
{\cal S}^s_{\vec q} &=& \frac{1}{\cosh r_q}
                \exp\left\{
                        -{a^s_{\vec q}}^\dagger(\eta_0){a^s_{-\vec
                        q}}^\dagger(\eta_0) e^{2i\phi_q}\tanh r_q
                    \right\}
\nonumber \\
        &{}&
        \exp\left\{-\left[{a^s_{\vec q}}^\dagger(\eta_0) a^s_{\vec q}(\eta_0)
                +
                {a^s_{-\vec q}}^\dagger(\eta_0) a^s_{-\vec q}(\eta_0)
                \right]\ln\cosh r_q
                \right\}
\nonumber \\
        &{}&
        \exp\left\{a^s_{\vec q}(\eta_0){a^s_{-\vec
	q}}(\eta_0)e^{-2i\phi_q}\tanh r_q
                    \right\}.
\label{e20}
\end{eqnarray}
Then
\begin{equation}
{{\cal S}^s_{\vec q}}^\dagger\vert0\rangle = \frac{1}{\cosh r_q}
        \sum^\infty_{n=0}(e^{2i\phi_q}\tanh r_q)^n\vert n, n\rangle\>,
\label{e21}
\end{equation}
so that $\langle0\vert\cos\varphi^s_{\vec q}(\eta)\vert0\rangle = 0$.
Similarly, $\langle0\vert\sin\varphi^s_{\vec q}(\eta)\vert0\rangle = 0$.
The average phase of any one mode vanishes, as expected for
randomly distributed phases. This, however, does not tell us whether
or not this random distribution is ``classical''. To do so,
we calculate the rms fluctuation in the phase
\begin{equation}
(\Delta\cos\varphi^s_{\vec q})^2 = \langle0\vert(\cos\varphi_{\vec
	q}^s)^2\vert0\rangle = \frac{1}{2} -
	\frac{1}{4}\frac{1}{\cosh^2 r_q}\>,
\label{e22}
\end{equation}
with $\langle0\vert(\cos\varphi_{\vec q}^s)^2\vert0\rangle =
\langle0\vert(\sin\varphi_{\vec q}^s)^2\vert0\rangle$.

In eq.~$(\ref{e22})$ we can see a deviation from the classical
stochastic behavior. Suppose that the phase of this mode is classical
in nature and can be described by a random, stochastic behavior.
Denoting the phase of this mode by $\theta_q$, then because
everything are c-numbers, $\cos^2\theta_q +\sin^2\theta_q =1$. If
the phase of this mode is random, we would expect the stochastic
average $\langle\cos^2\theta_q\rangle_{sto} =
\langle\sin^2\theta_q\rangle_{sto}$. Consequently, for a classical,
random stochastic distribution of phase, one would expect
$\langle\cos^2\theta_q\rangle_{sto}=1/2$.

Since, however, $(\cos\varphi^s_{\vec q})^2$ and
$(\sin\varphi^s_{\vec q})^2$ are {\it operators}, their sum need
not add up to unity. As such, for a random distribution their
expectation value (average) need not be $1/2$, as it is for the
classical stochastic distribution. In fact, any deviation from
$1/2$ is a sign of the quantum nature of the mode. As we can see from
eq.~$(\ref{e22})$, the quantum nature of the mode is always present,
but becomes progressively smaller for large $r_p$. Modes just
entering the horizon at the present day have a $r_q\sim 120$ and for
these modes $\langle 0 \vert (\cos\varphi^s_{\vec q})^2\vert 0
\rangle \approx 1/2$. They are therefore essentially classical in
nature and using a random, stochastic distribution to describe their
phase would be correct. Notice, however, that for $r_q \sim 1$,
deviation from the classical behavior becomes pronounced and these
modes are essentially quantum mechanical in nature.

Of course, the absolute phase of any one mode is irrelevant. What is
more interest is the relative phases between modes. Given the phase
operator for any one mode, we follow $\cite{CarNieto}$ and define
the phase sum and difference operators between any two modes as
\begin{eqnarray}
\sin(\varphi^s_{\vec p} \pm \varphi^t_{\vec q})
	&\equiv& \sin\varphi^s_{\vec p}\cos\varphi^t_{\vec q}
		\pm
		\cos\varphi^s_{\vec p}\sin\varphi^t_{\vec q},
\nonumber \\
\cos(\varphi^s_{\vec p} \pm \varphi^t_{\vec q})
	&\equiv& \cos\varphi^s_{\vec p}\cos\varphi^t_{\vec q}
		\mp
		\sin\varphi^s_{\vec p}\sin\varphi^t_{\vec q}.
\label{e23}
\end{eqnarray}
Notice that in the limit $\vec p\to\vec q$, $t=s$ the sine difference
operator {\it does not vanish\/} since $[\cos\varphi^t_{\vec
q},\sin\varphi^t_{\vec q}]\ne0$. This once again underscores the
fact that we are dealing with operators and not functions.

It is then straightforward to show that for all $t$, $s$, and $\vec
p\ne- \vec q$,
\begin{equation}
\langle0\vert\sin(\varphi^s_{\vec p} \pm \varphi^t_{\vec
q})\vert 0\rangle = 0\>,
\qquad\qquad\langle0\vert\cos(\varphi^s_{\vec p} \pm
\varphi^t_{\vec q})\vert 0\rangle = 0\>,
\label{e24}
\end{equation}
as expected since the average phase of each mode vanishes. What is of
more interest is the expectation value of the squares of these
operators
\begin{eqnarray}
\langle0\vert(\sin(\varphi^s_{\vec p} - \varphi^t_{\vec
q}))^2 \vert 0\rangle
	&=& \langle0\vert(\cos(\varphi^s_{\vec p} -
\varphi^t_{\vec q}))^2\vert 0\rangle = \frac{1}{4}(\tanh^2 r_p +
\tanh^2 r_q)\>,
\nonumber \\
\langle0\vert(\sin(\varphi^s_{\vec p} + \varphi^t_{\vec
q}))^2 \vert 0\rangle
	&=& \langle0\vert(\cos(\varphi^s_{\vec p} +
\varphi^t_{\vec q}))^2\vert 0\rangle = \frac{1}{4}(1+\tanh^2
r_p\tanh^2 r_q)\>.
\nonumber \\
&{}&
\label{e25}
\end{eqnarray}
Once again in the limit of large $r_p$ and $r_q$, these results go
over to what one expects for a classical stochastic distribution of
the phases. Consequently, we find that as long as $\vec p\ne - \vec q$
the phase difference and sum between modes are completely random and
in the limit of large $r_p$ and $r_q$, can be accurately approximated
as classical stochastic distribution of the phases.

When $t=s$ and $\vec p = -\vec q$ the situation changes quite
dramatically, however. First, we find that
\begin{equation}
\langle0\vert\sin(\varphi^s_{\vec p} - \varphi^s_{-\vec
p}) \vert 0\rangle = 0,
\qquad\qquad\langle0\vert\cos(\varphi^s_{\vec p} -
\varphi^s_{-\vec p})\vert 0\rangle = 0,
\label{e26}
\end{equation}
as expected. But now
\begin{eqnarray}
\langle0\vert\sin(\varphi^s_{\vec p} + \varphi^s_{-\vec
p}) \vert 0\rangle &=& \sin2\phi_p\tanh r_p\>,
\nonumber \\
\qquad\qquad\langle0\vert\cos(\varphi^s_{\vec p} +
\varphi^s_{-\vec p}\vert 0\rangle &=& \cos2\phi_p\tanh r_p\>.
\label{e27}
\end{eqnarray}
Moreover,
\begin{equation}
\langle0\vert(\sin(\varphi^s_{\vec p} - \varphi^s_{-\vec
p}))^2 \vert 0\rangle = \langle0\vert(\cos(\varphi^s_{\vec p} -
\varphi^s_{-\vec p}))^2\vert 0\rangle = \frac{1}{2}\tanh^2r_p\>,
\label{e28}
\end{equation}
while
\begin{eqnarray}
\langle0\vert(\sin(\varphi^s_{\vec p} + \varphi^s_{-\vec
p}))^2 \vert 0\rangle &=&
\frac{1}{4}\left(1-\tanh^2 r_p+ 4\tanh^2 r_p\sin^2(2\phi_p)\right) \>,
\nonumber \\
\langle0\vert(\cos(\varphi^s_{\vec p} +
\varphi^s_{-\vec p})^2\vert 0\rangle &=&
\frac{1}{4}\left(1-\tanh^2 r_p+ 4\tanh^2 r_p\cos^2(2\phi_p)\right).
\label{e29}
\end{eqnarray}
Once again, in the limit of large $r_p$, we find that
$(\Delta\sin(\varphi^s_{\vec p} - \varphi^s_{-\vec p}))^2=1/2$,
meaning that phase difference between the two modes is completely
random. The fluctuation in the phase sum $\Delta\sin(\varphi^s_{\vec
p} + \varphi^s_{-\vec p})=0$, however, and the two modes have a
definite {\it phase sum\/}. This is a manifestation of two mode phase
locking first calculated by $\cite{BarPegg}$ using distribution
function methods.

We thus see that although the phase differences between modes
are uncorrelated and completely random, the phase sum between the
$\vec p$ and $-\vec p$ modes with the same polarization is highly
correlated with an average value of $2\phi_p$ and essentially no
fluctuation whatsoever at large $r_p$. This correlation was also
founded by Grishchuk and Sidorov and was interpreted by them as the
formation of a classical standing wave. Note, however, that this
phenomenon is inherently quantum mechanical in nature and it is
impossible to explain both eqs.~$(\ref{e28})$ and ~$(\ref{e29})$
using classical stochastic arguments for the following reasons.

Suppose that we wish to explain the results of the phase sum and
difference analysis for $\vec p = -\vec q$ in the large $r_p$ limit
using classical stochastic arguments. Then eq.~$(\ref{e27})$
implies that the phase distribution of both the $\vec
p$ and $-\vec p$ modes are peaked at $\phi_p$. That there is a width
to this distribution can be seen in eq.~$(\ref{e28})$ and in fact we
see that there must be a randomly distributed background noise below
$\phi_p$. If, however, this background noise is present, then one
would not expect eq.~$(\ref{e29})$ to hold, since although we would
expect the average phase sum to be $2\phi_p$, we still expect the
noise to be present and would not expect the rms fluctuation in the
phase sum to vanish. Classically, it should be $1/2$ once more.
Consequently, the results of the phase sum and difference analysis
cannot be explained using classical methods.

The phase locking of the $\vec p$ and $-\vec p$ modes also suggests the
following method to measure the thermal history of the universe.
Notice that for large $r_p$, which correspond to most physically
relevant modes, the phase sum of the two modes is $2\phi_p$ with
essential no fluctuation whatsoever. From eq.~$(\ref{e6})$, we see
that the value of $\phi_p$ depends explicitly on the scale factor $R$
which is in turn determined by the thermal history of
the universe. Consequently, we propose to fix a specific direction
along the celestial sphere and measure the phase sum of primordial GW
with various momentum. This will determine each $\phi_p$. From
eq.~$(\ref{e13})$, $r_p$ can be infered by measuring the amplitude of
the wave, although from eq.~$(\ref{e18})$ the fluctuation in this
amplitude is expected to be large. Through these two measurements the thermal
history along that direction can be extracted. Next, fix the magnitude of
$\vec p$ and measure the phase sum $=2\phi_p$ for a series of angles
on the celestial sphere. If the universe was truly isotropic
throughout its history, then $\phi_p$ should be independent of angle.
If not, then variation in $\phi_p$ will be a measure of the
anisotropy of the universe. In either case, a complete determination
of $\phi_p$ and $r_p$ will be useful for probing the very early
universe.

How these experiments would be done is not as yet known. Not only hasn't
any GW been detected yet (not to mention a primordial one), this
experiment would also require measuring the quantum mechanical phase
sum of two modes. It has only been very recently that the phase operator for
optical waves has been measured experimentally $\cite{Expt}$,
and no experimental measurement of phase locking has yet been found
even for optical waves. Consequently, we would expect such
experiments, if they can be done at all, can only be accomplished in
the far, far future. Nevertheless, they have the potential to provide
a very clean and direct measurement of the history of the universe.

To conclude, we have in this paper completed a detailed quantum
mechanical analysis of the properties of GW arising from fluctuations
in the de Sitter vaccuum of the universe. We find that the present
day charactorization of these waves as classical GW with a classical,
stochastic distribution in the phase to be somewhat naive. First,
while one can identify the rms fluctuation of each mode of the
graviton field $h_{ij}$ as a classical amplitude, the fluctuation of
this `amplitude' is very large for large $r_p$, the most
physically relevant ones. Next, although the average phase of each mode
seperately is randomly distributed and can be well approximated as a
classical stochastic distribution, there are non-classical, strong
correlations between the phases of different $\vec p$. There is a
definite value of the phase sum of the $\vec p$ and $-\vec p$ modes,
although the phase difference of these modes, as it
is between any two modes, is completely random and stochastically
distributed. This has been interpreted by Grishchuk and Sidorov as
being due to the formation of standing waves in the universe.

We now can understand the limitations of the usual classical
treatment of the relic GW. In this treatment, the time evolution of
the amplitude of classical GW generated from inflation is obtained
from a two-point statistical average of a stochastic ensemble of
classical gravitational fields which is then matched to the quantum
mechanical two-point function of the gravitational field
$\cite{Abbo}$. The subsequent evolution of the classical waves is
then described by the classical wave propagation in the expanding
universe. This constitutes the so-called stochastic classical
background of relic GW with randomly distributed phases. Intuitively,
we say that during inflationary expansion fluctuations in the
gravitational field will be red-shifted out of the horizon, after
which they freeze and remain at a constant amplitude. Much later when
they re-enter the horizon during either the radiation- or
matter-dominated era, they will appear as classical oscillating GW.

There are, however, two different phases which must be considered
if one wishes to view the system in this way. The first is the {\it
temporal\/} phase of oscillation of waves. This is precisely fixed in
inflationary cosmologies.  Once a frozen mode starts to oscillating
as a classical wave at the time it crosses the horizon and re-enters
the universe, the temporal phase is completely determined by the
classical equation of motion. However, inflation does not predict the
{\it location\/} of the nodes of these frozen modes since quantum
fluctuations assign equal probability to modes which differ only by a
spatial translation. Herein lies the other phase which must be
considered: that due to the spatial part of the GW. It is this phase
which is completely random. Consequently, in this classical treatment
the overall phase of GW are randomly distributed.

Based on our results of the statistical properties of the phases of GW, the
above classical description has to be amended. Since the phases
of different modes are almost uncorrelated, in many aspects the relic
GW can be well as a stochastic classical background radiation.
Phase-sum locking, however, is still present and is a strong
indication of its inherent quantum character of these GW.
Moreover, this seperation of the phase of each mode into spatial and
temporal parts is completely artificial and araises only from the
desire to use a classical description and interpretation of the
quantum two-point function. From the experience of calculating the
vacuum expectation values of the phase operators, we can hardly
separate the temporal phase from the spatial phase. This is not
surprising, though, since there is only one single overall phase in the
quantum mechanical treatment. Consequently, although the classical
approach advocated in $\cite{Abbo}$ is quite physical and is, as we
have seen, a good approximation of an inherently quantum mechanical
phenomenon in many instances, it can not fully describe the
properties of the relic GW. Our results suggest that one should
instead use the quantum mechanical approach to deal with these GW.

\begin{center}
{\bf Acknowledgements}
\end{center}

This work was supported in part by the R.O.C. NSC Grant Nos.
NSC82-0208-M-001-131-T and NSC83-0208-M001-69.

\end{document}